# Colossal Magnetoresistance Study in Nanophasic $La_{0.7}Ca_{0.3}MnO_3$ Manganite

U.K. Goutam[a], P.K. Siwach[a], H.K. Singh[b], R.S. Tiwari[a] and O.N. Srivastava[a#]

[a]Department of Physics, Banaras Hindu University, Varanasi-221005, India
[b]National Physical Laboratory, Dr K S Krishnan Marg, New Delhi-110012, India

**Abstract**

In this study we report the effect of sintering temperature on the low field magnetotransport properties of $La_{0.7}Ca_{0.3}MnO_3$ manganite synthesized through polymeric precursor route. The $La_{0.7}Ca_{0.3}MnO_3$ has been sintered at $600^oC$ (S6), $700^oC$ (S7), $800^oC$ (S8), $900^oC$ (S9) and $1000^oC$ (S10). XRD confirms that phase formation starts at $600^oC$. All the samples are single phasic having orthorhombic unit cell. The lattice parameters decrease on lowering the sintering temperature ($T_S$). The crystallite as well as particle size also show strong dependence on the sintering temperature. All the samples possess insulator-metal ($T_{IM}$) as well as paramagnetic-ferromagnetic ($T_C$) transitions. The $T_C$ shows a small variation [274K (S10) to 256K (S6)] as a function of sintering temperature whereas the $T_{IM}$ goes down from 267K (S10) to 138K (S6), a strong decrease of 129K. This $T_C$-$T_{IM}$ discrepancy is due to the fact that whereas the former is an intrinsic characteristic, the latter depends strongly on the extrinsic factors e.g. synthesis conditions grain boundaries and associated disorders. For all the samples MR shows strong dependence on $T_S$. The MR increases on lowering the temperature as well as on increasing the field with the occurrence of an intrinsic contribution around $T_C$. These variations of MR for all the samples have been explained in terms of the microstructural variations and spin-polarized tunneling at low temperatures.

[#]Corresponding Author
Email: hepons@yahoo.com

**Introduction**

The colossal magnetoresistance (CMR) in hole doped manganese oxides widely known as manganites with formula $L_{1-x} A_x MnO_3$ (where L is trivalent rare earth and A is divalent alkaline earth ion) has been intensively studied over the last decade. The interest in these materials is due to their unusual magnetic and electronic properties along with possible technological applications in magnetic information storage and faster reading devices [1-3]. The complexities involved in their magnetic, electronic and structural properties leads to a host of phenomena that are yet poorly understood. Some of these are insulator-metal, paramagnetic-ferromagnetic transitions, charge/orbital ordering and nanoscale phase separation. The inter-relationship between these makes them fascinating for the basic research as well [4-6].

The most prominent theory capable of explaining the major results of CMR is Double Exchange (DE) mechanism [7-9]. In this, there is exchange of electrons from neighbouring $Mn^{3+}$ to $Mn^{4+}$ ions through oxygen when their core spins are parallel and hopping is not favoured when they are anti-parallel. However, this model cannot properly explain all the details of observed CMR effect. Therefore, other theories have been developed, which besides DE mechanism also incorporate the Jahn Teller character of $Mn^{3+}$ ion by a variable electron-phonon coupling [10]. The concept of phase separation (PS) has recently emerged according to which the physics of manganites in the CMR regimes is dominated by inhomogenities in the form of co-existing competing phases [11]. These doped perovskite manganites show large magnetoresistance around $T_{IM}$ ($T_C$) at a high magnetic field, which restricts their applicability to hands on devices. However, it was further realized that manganite samples containing a large number of defects, particularly grain-boundaries, show a large low-field magnetoresistance (LFMR) generally absent in single crystals and epitaxial films. This LFMR or grain boundary MR (GBMR) is of prime importance for application and is due to spin-polarized tunneling between neighbouring grains [12].

During the last decade nanocrystalline form of various materials have drawn considerable attention because they typically exhibit physical and chemical properties that are distinct from their bulk counterparts. The physical properties of manganites are also expected to depend on material size due to both the nanoscale phase inhomogeneities inherent to these materials and additional surface effects. Hwang et al [12] pointed out that the large low field magnetoresistance (LFMR) of the polycrystalline samples is dominated by spin polarized tunneling between neighbouring grains. This is quite different from their single crystals and epitaxial films counterparts where the double exchange mechanism is prominent. The bulk samples of manganites were usually synthesized by the conventional ceramic methods that need higher sintering temperature and long annealing time to obtain homogenous composition and desired structure. These methods are not appropriate for many advanced applications, due to formation of large particles, agglomerates, poor homogeneity, undesirable phases, abnormal grain growth and an imprecise stoichiometric control of cations. However, the sol gel process has potential advantage over the other methods not only for achieving homogenous mixing of the components on the atomic scale, but also for the possibility of forming desired shapes which are of technological importance. Other advantages of the sol-gel route are lower processing temperatures, short annealing times, high purity of materials, good control of size and shape of the particles and particle size well below 100nm at the lower processing temperature. It may be opportune to mention that there are several reports on the synthesis of nanophasic manganites by sol gel based synthesis methods but none of them seems to have carried out extensive studies on the effect of sintering temperature on sol-gel synthesized manganites in relation to low field magneto-transport properties [13-22]. In this work, we report the synthesis of $La_{0.7}Ca_{0.3}MnO_3$ perovskite manganite by low temperature polymeric precursor route and studied the effect of sintering temperature on microstructure and low field magneto-transport properties.

**Experimental**

We have adopted the sol-gel based polymeric precursor route to synthesize $La_{0.7}Ca_{0.3}MnO_3$ (LCMO) samples having nano size particles at a significantly lower sintering temperature compared to the conventional solid-state procedure. In this technique, aqueous solution of high purity $La(NO_3)_3.6H_2O$, $Ca(NO_3)_2.4H_2O$ and $Mn(NO_3)_2.4H_2O$ have been taken in the desired stoichiometric proportions. An equal amount of ethylene glycol has been added to this solution with continuous stirring. This solution is then heated on a hot plate at a temperature of ~ 100-140$^0$C till a dry thick brown colour sol is formed. At this temperature ethylene glycol polymerizes into polyethylene glycol, which disperses the cations homogeneously forming cation-polymer network. This has been further decomposed in an oven at a temperature of ~ 300$^0$C to get polymeric precursor in the form of black resin like material. The polymeric precursor thus obtained is then sintered at different temperatures ranging from 500 to 1000$^0$C. Several time periods ranging between 2 to 6 hrs were employed. It was found that sintering for ~4 hrs gave optimum results for all temperatures. Phase pure completely crystalline samples have been obtained at the temperature as low as 600$^0$C. The    LCMO samples sintered at 600°C, 700°C, 800°C, 900°C and 1000°C will hereafter be referred to as S6, S7, S8, S9 and S10 respectively.

All the synthesized samples have been subjected to gross structural characterizations using powder X-ray diffractometer [XRD, Philips PW 1710] using Cu K$\alpha$ radiation at room temperature and microstructural characterization by scanning electron microscopic technique [SEM, Philips XL20]. The magnetotransport measurements have been done by standard dc four-probe technique in the temperature range of 300-80K and in applied magnetic field in the range of 1.0T. The magnetic characterizations have been carried out by ac susceptibility measurements.

**Results and discussion**

**a. Structural and Microstructural Characterizations:**

Sintering temperature is one of the key factors that influence the crystallization and microstructure of the perovskite $La_{0.7}Ca_{0.3}MnO_3$ (LCMO) samples, which in turn affect the magneto-transport properties. The crystallinity and phase analysis of all the synthesized samples (S6, S7, S8, S9 and S 10) were determined by the powder X-ray diffraction and the corresponding pattern are shown in Fig.1a. All the samples are orthorhombic and single phasic without any detectable secondary phase. In the present polymeric precursor method, the characteristic perovskite phase formation starts at a significant low temperature of 600°C as compared to other conventional methods. The lattice parameters (orthorhombic unit cell parameters a, b, c and the unit cell volume V= abc) decrease in a systematic fashion with increasing sintering temperature as shown in Table1. The intensity of the X-ray peaks for the LCMO perovskite phase increases as the sintering temperature increases from 600°C to 1000°C indicating that the crystallinity of $La_{0.7}Ca_{0.3}MnO_3$ becomes better with higher sintering temperature. Fig. 1b shows the (200) reflection of S6, S7, S8, S9 and S10 samples. It is clear from inset that as the sintering temperature increases there is a decrease in the full width at half maximum (FWHM) and hence the crystallite size increases. The average crystallite sizes of the samples are obtained by the X-ray line width using Scherrer formula P.S. $\simeq k\lambda /\beta \cos\theta$; where $k \simeq 0.89$ is shape factor, $\lambda$ is wavelength of X-rays, $\beta$ is the difference of width of the half maximum of the peaks between the sample and the standard silicon used to calibrate the intrinsic width associated with the equipment and $\theta$ is the angle of diffraction. The average crystallite size of a sample sintered at 700°C (S7) is calculated to be ~32nm. The same increases to ~ 35nm, 40nm and 62 nm for sample sintered at 800°C (S8), 900°C (S9) and 1000°C (S10) respectively. The crystallite sizes, lattice parameters and unit cell volumes obtained for the different samples are listed in Table 1. Fig. 2 (a-d) shows the representative images elucidation surface morphology for the samples S10, S9, S8 and S7 respectively employing scanning electron microscope in secondary electron imaging mode. SEM observation reveals that there is a distribution of particle size for all samples and as the sintering temperature increases, the particle size increases and the porosity decreases. The highest temperature (1000°C) sintered sample (S10) has well connected particles whereas as we go down to lower temperature sintered sample (S7), the particle connectivity becomes poor. The average particle size is ~ 50, 75, 125 and 200 nm respectively for samples S7, S8, S9 and S10. Fig.3 shows the variation of crystallite sizes (CS) obtained from the width of the X-ray diffraction peaks and the particle sizes (PS) obtained from SEM. Both crystallite as well as particle size increase as the sintering temperature is increased due to congregation effect. However, it has been observed that there is a difference between CS and PS at all sintering temperature and is more pronounced at higher sintering temperature. For example, CS = 32 nm and PS = 50 nm for S7 and for S10 it is ~ 62 and ~ 200nm respectively. This difference is due to the fact that particles are composed of several crystallites, probably due to the internal stress or defects in the structure [21].

**b. Electrical Transport Characteristics:**

The paramagnetic-ferromagnetic transition temperature ($T_C$) has been measured by measuring the ac susceptibility ($\chi$) in the temperature range of 300-80K. The variation of $\chi$ with temperature for the samples S6, S7, S8, S9 and S10 is shown in Fig.4, which depicts a ferromagnetic ordering transition for all the samples. We have observed only a slight variation in $T_C$ for the samples sintered at different temperatures, which have been examined by the peaks in ($d\chi/dT$). The values of $T_C$ are given in Table 2. As is clear from this table $T_C$ shows an increase from 256K for the sample S6 to 274K for the sample S10. It has also been observed that as the sintering temperature decreases the width of transition broadens which suggests that at low sintering temperatures grains are loosely connected as also visible in the scanning electron micrograph shown in Fig.2. One of the main advantages of polymeric precursor method, as we have stated earlier, is enhancement in $T_C$ as compared to standard solid-state method, which is seen here up to ~30K. Also Fig.4 indicates that the magnetization of the samples increases as the sintering temperature increases which is same as found in earlier results [22].

The dc resistivity ($\rho$) was measured in the temperature range of 300-80 K by four probe technique with and without magnetic field and the data for resistivity without applied magnetic field for samples S6, S7, S8, S9 and S10 are plotted in Fig.5. At room temperature (~300 K), resistivity values are ~ 309.15, 221.48, 89.18, 19.25 and 2.42 Ω-cm, respectively for S6, S7, S8, S9 and S10 samples. Thus even at room temperature large enhancement, by more than two orders of magnitude, in the resistivity is observed for S6 as compared to S10 as a consequence of lower $T_S$ and hence smaller particle size. This increase in resistivity is believed to be caused mainly due to enhanced scattering of the charge carriers by the higher density of magnetic disorder in GBs at smaller particle size. On increasing $T_S$ the particle size increases leading to decrease in the GBs and the associated disorder. This results in decrease in scattering of the carriers expressed by a decrease in the resistivity. All the samples show an increase in the resistivity on lowering temperature and at a characteristic temperature, which is lower than the corresponding $T_C$, an insulator ($d\rho/dT < 0$) to metal ($d\rho/dT > 0$) like transition is observed. The measured characteristic insulator-metal transition temperatures ($T_{IM}$) are ~267, 240, 210, 179 and 138K for S10, S9, S8, S7 and S6 respectively. The corresponding peak resistivity values are ~ 3.91, 33.9, 241.55, 1013.51 and 3748.27 Ω-cm respectively. The sol-gel prepared samples show a large difference between $T_C$ and $T_{IM}$ and the difference increases as we lower the sintering temperature. This variation of $T_C$ and $T_{IM}$ with the sintering temperature is plotted in Fig.6. The large difference in the $T_C$ and $T_{IM}$ for all the LCMO samples is thought to be due to the existence of the disorder and is in fact a common feature of the polycrystalline manganites [23]. The $T_C$ being an intrinsic characteristic does not show significant change as a function of the sintering temperature. On the other hand $T_{IM}$ is an extrinsic property that strongly depends on the synthesis conditions and microstructure (e.g. grain boundary density). Thus the $T_{IM}$ goes down by 129K on lowering the sintering temperature from 1000°C to 600°C whereas Tc undergoes a small decrease with a concomitant transition broadening. The strong suppression of the $T_{IM}$ as compared to $T_C$ is caused by the induced disorders and also by the increase in the non-magnetic phase fraction, which is due to enhanced grain boundary densities as a consequence of lower sintering temperature. This also causes the increase in the carrier scattering leading to a corresponding enhancement in the resistivity. Thus lowering of sintering temperature reduces the metallic transition temperature and hence the concomitant increase in resistivity. When a magnetic field is applied, the FM clusters grow in size and the interfacial Mn spin disorder is suppressed resulting in the improved connectivity and consequently a decrease in the resistivity has been observed.

**c. Magneto-transport Characteristics:**

The property that is of prime importance for these doped perovskite manganites is their magnetoresistance. The magnetoresistance (MR) is calculated by the formula MR (%) = $[(\rho_0-\rho_H)/ \rho_0]$ x 100; where $\rho_0$ and $\rho_H$ are the resistivity measured at H = 0 and H respectively. The temperature dependence of MR for S7, S8, S9 and S10 samples measured in the range 80-300 K at 3 kG and 10 kG are shown in Fig.7. All the samples show a sequential increase in low temperature MR with decreasing sintering temperature. The appearance of peak in the MR-T curve around $T_C$ depicts that in all the samples there is a contribution of the intrinsic component of MR, which arises due to the double exchange mechanism around $T_C$. However, around $T_C$ the peak in the MR-T curve of the sample S10 is significantly higher at all applied magnetic fields in comparison to other samples. The magnitude of MR peak around $T_C$ decreases as we lower the sintering temperatures as well as we reduce the magnitude of the applied magnetic field (Fig.7).

The peak MR values are ~13.07 and 10.34 % at 10 kG applied field for samples S10 and S9 whereas for sample S8 and S7 there is a hump in the MR variation around $T_C$. At 80 K, the MR values are measured to be ~12.87, 13.66, 13.85 and 15.46 for S10, S9, S8 and S7 respectively at the field of 3 kG. Thus, decreasing crystallite/grain size leads to the enhancement in low field MR at lower temperatures while the MR in the higher temperature regime is suppressed. The disappearance of the high temperature MR can be explained by weakening of the double exchange (DE) mechanism around the respective PM-FM transition temperatures due to decrease in particle size which results from low sintering temperature. It has been found that all the samples are showing significant MR at low field. The low field MR (LFMR) increases as the sintering temperature and hence particle size decreases. This is consistent with previous studies [21,22].

The magnetic field dependence of the low field MR at 80 K and 150K for all the samples are given in Figure 8. The low field MR of all the samples is observed to increase with increasing magnetic field. The MR-H curve shows two different slopes, the one below H ~ 1.5 kG is steeper while the other above H ~ 3 kG is rather weak. As the sintering temperature is lowered the MR increases and the slope of the MR-H curve in both the temperature regimes becomes steeper. For example in the S10 sample, the low field MR at H = 1.5 kG and 12 kG is measured to be ~ 6.37 % and 16.4 %, respectively at 150K whereas the same increases to ~ 12.16 % and 22.56% at 80K. The LFMR values at 150K and 80K for all the LCMO samples are listed in Table 2. It should also be noted that the variation of MR does not show any saturation in MR even up to 12 kG fields. The enhanced slope of the 80K MR-H curves in the low field regime as compared to those taken at 150K suggests that in the lower temperature regime spin polarized transport may the MR causing mechanism. These studies clearly depict that polymeric precursor technique is an effective method to enhance ferromagnetic ordering temperature along with enhancement of low field magnetoresistance.

**CONCLUSION**

In summary, we have studied the effect of sintering temperature on microstructure and low field magneto-transport properties of polycrystalline nanophasic $La_{0.7}Ca_{0.3}MnO_3$ (LCMO), which have been successfully synthesized by polymeric precursor route. All the LCMO samples are single phasic and have orthorhombic unit cell. The importance of the polymeric precursor route is that the perovskite phase formation takes place at relatively low temperature of ~600$^o$C. The lattice parameter decreases in a systematic way as we lower the sintering temperature (Table-1). Microstructure reveals that particle size also decreases from ~200 nm (S10) to ~50 nm (S7). The paramagnetic-ferromagnetic and insulator -metal transitions have been observed in all the samples. Both the transition temperatures $T_C$ and $T_{IM}$ shift towards lower temperatures as the particle size decreases (due to lowering of sintering temperature). There is only a slight decrease in $T_C$ (from 274K to 256K) whereas $T_{IM}$ decreases considerably from 267K to 138K as the sintering temperature is lowered from 1000 to 700 °C. Synthesis of LCMO by polymeric precursor route also resulted in considerable enhancement of $T_C$ up to 274K. Also, we have observed a huge low field MR along with intrinsic MR. It has been found that low field MR increases as the sintering temperature (particle size) decreases but at the same time peak (intrinsic) MR decreases. This enhanced LFMR for small size particles is due to increased spin polarized tunneling behavior at lower temperatures. A detailed study on magnetic behaviour and magnetoresistance properties with particles size is in progress and results will be forthcoming.

**Acknowledgement**

This work was supported financially by UGC and CSIR, New Delhi. One of the authors (PKS) acknowledges CSIR, New Delhi for the award of SRF. Authors are also thankful to Professors A R Verma, C N R Rao, T V Ramakrishnan, Vikram Kumar, S B Ogale, A K Raychaudhari and Dr Kishan Lal for valuable discussions.

**References**

1. Haghiri-Gosnet A-M and Renard J-P 2003 *J. Phys. D: Appl. Phys.* **36** R127-R150

2. Levy P M, 1994 *Solid State Phys.* **47** 367

3. Daughton J M, 1999 *J. Magn. Magn. Mater.* **192** 334

4. Tokura Y (ed.) 2000 *Colossal Magnetoresistive Oxides,* (Gordon and Breach, London)

5. Rao C N R and Raveau B (ed.) 1998 *Colossal Magnetoresistance, Charge Ordering and Related Properties of Manganese Oxides* (World Scientific: Singapore)

6. Dagotto E (ed.) 2002 *Nanoscale Phase Separation and Colossal Magnetoresistance* (Springer-Verlag, Berlin, New York)

7. Zener C 1951 *Phys. Rev.* **82** 403

8. Anderson P W and Hasegawa H 1955 *Phys. Rev.* **100** 675-81

9. de Gennes P G 1960 *Phys. Rev. B* **118** 141.

10. Millis A J, Littlewood P B and Shraiman B 1995 *Phys. Rev. Lett.* **74** 5144-7

11. Dagotto E, Hotta T and Moreo A 2001 *Phys. Rep.* **344** 1-153

12. Hwang H Y, Cheong S-W, Ong N P and Batlogg B 1996 *Phys. Rev. Lett.* **75** 2041-4

13. Sánchez R D, Rivas J, Vázquez-Vázquez C, López-Quintela M A, Causa M T, Tovar M and Oseroff S B 1996 *Appl. Phys. Lett.* **68(11)** 134-7

14. Lisboa-Filho P N, Mombru A W, Pardo H, Ortiz W A and Leite E R, 2003 *J. Phys. Chem. Solids* **64** 583

15. Rivas J, Hueso L E, Fondado A, Rivadullo F and Lopez-Quintela M A, 2000 *J. Magn. Magn. Mater.* **21** 57

16. Pandya Dinesh K, Kashyap Subhash C and Pattanaik Gyana R 2001 *Journal of Alloys and Compounds* **326** 255–259

17. Vertruyen B, Rulmont A, Cloots R, Ausloos M, Dorbolo S and Vanderbemden P 2002 *Mater. Lett.* **57** 598–603

18. Lisboa-Filho P N, Mombru A W, Pardo H, Leite E R and Ortiz W A 2004 *Solid State Communications* **130** 31–36

19. Mahesh R, Mahendiran R, Raychaudhuri A K and Rao C N R 1996 *Appl. Phys. Lett.* **68** 2291.

20. Shankar K S, Kar S, Subbanna G N and Raychaudhuri A K 2004 *Solid State Communications* **129** 479–483

21. Vazquez-vazquez C, Blanco M C, Lopez-quintela M A, Sanchez R D, Rivas J and Oseroff S B 1998 *J. Mater. Chem.,* **8(4)** 991-1000

22. López-Quintela M A, Hueso L E, Rivas J and Rivadulla F 2003 *Nanotechnology* **14** 212-219

23. Siwach P K, Singh D P, Singh H K, Khare N, Singh A K and Srivastava O N 2003 *J. Phys. D: Appl. Phys.* **36** 1361-136.

**FIGURE CAPTIONS**

Fig.1 (a) Powder X-Ray Diffraction pattern of the as synthesized samples sintered at 600 (S6), 700 (S7), 800 (S8), 900 (S9) and 1000°C (S10). (b) Shows the width of the peaks for different sintered samples.

Fig.2 SEM micrographs of the samples (a) S10, (b) S9, (c) S8, and (d) S7 revealing surface morphology and particle size distribution.

Fig.3 Variation of crystallite sizes (by XRD) and grain sizes (by SEM) with sintering temperature for LCMO samples.

Fig.4 AC susceptibility vs. temperature curves depicting paramagnetic-ferromagnetic transition temperature ($T_C$).

Fig.5 Temperature dependence of resistivity ($\rho$) of the samples sintered at different temperatures.

Fig.6 Variation of insulator-metal transition temperature ($T_{IM}$) and paramagnetic-ferromagnetic transition temperature ($T_C$) with sintering temperature ($T_S$).

Fig.7 Magnetoresistance (MR%) as a function of temperature for applied magnetic field of 3 kG and 10 kG for the samples sintered at different temperatures.

Fig.8 Magnetoresistance (MR%) as a function of magnetic field (kG) at 150K and 80K for the samples sintered at different temperatures.

**TABLE CAPTIONS**

Table1 Lattice parameters, cell volumes, crystallite sizes (XRD) and particle sizes (SEM) of the samples sintered at different temperatures ($T_S$).

Table2 Paramagnetic-ferromagnetic transition temperature ($T_C$), insulator-metal transition temperature ($T_{IM}$) and low field magnetoresistance (LFMR) at the temperatures 80K and 150K for the field of 10 kG for all the samples.

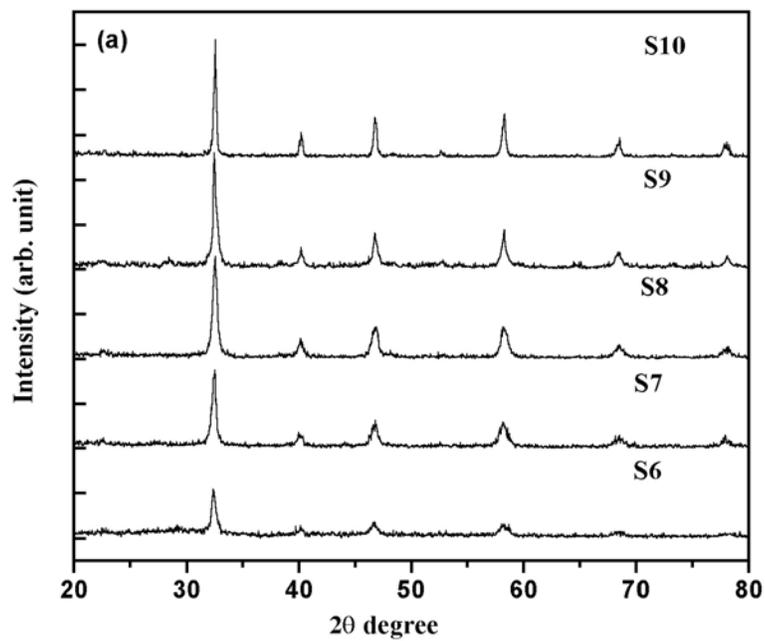

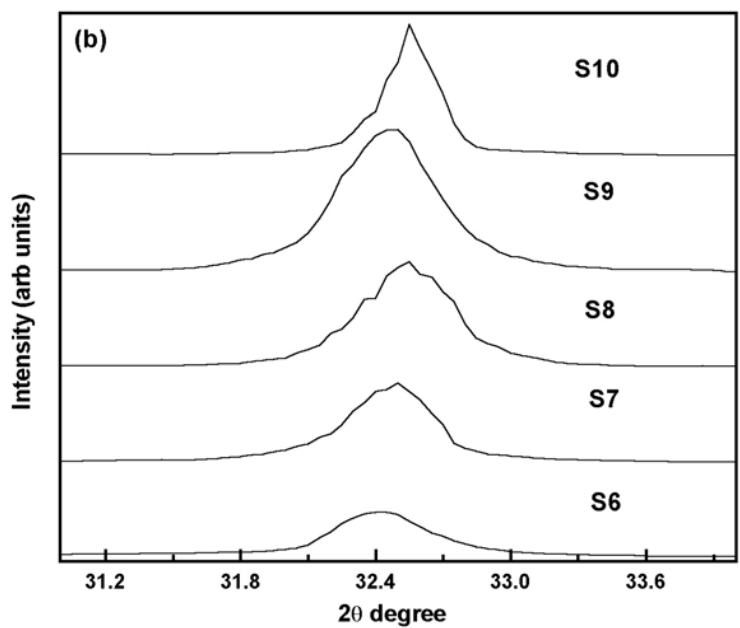

**Figure 1**

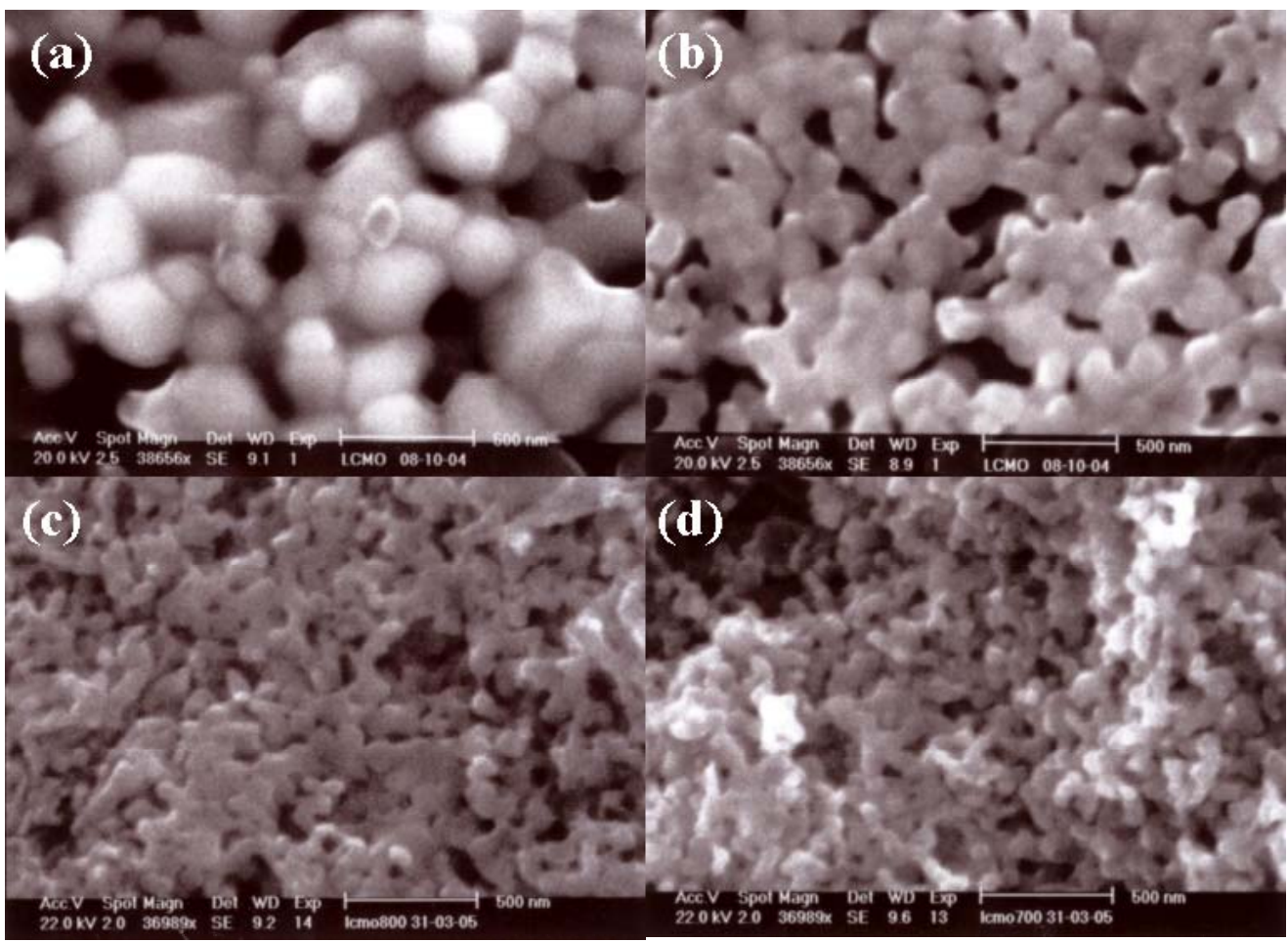

**Figure 2**

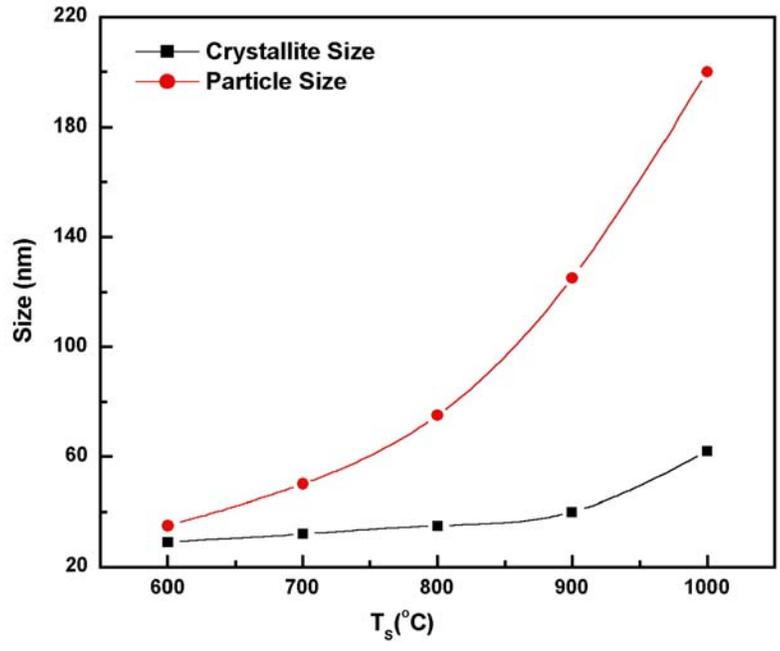

**Figure 3**

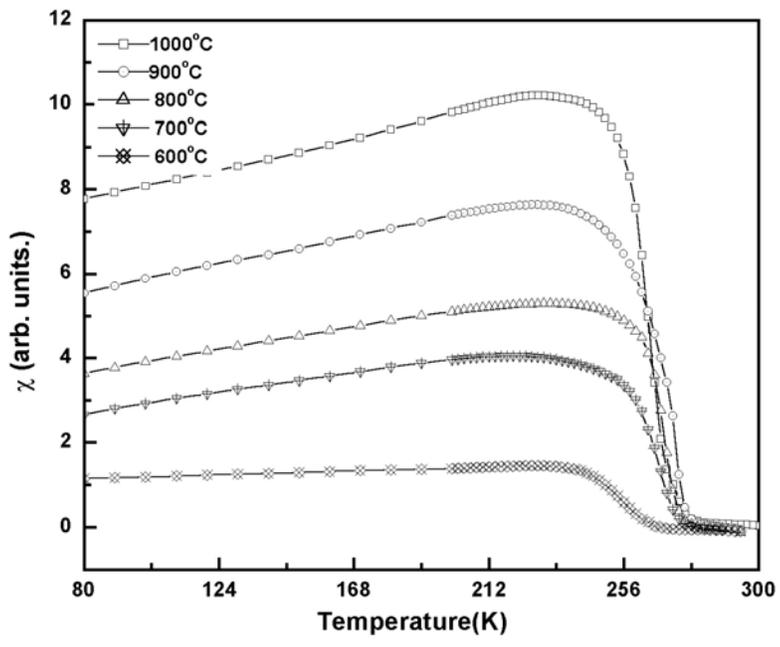

**Figure 4**

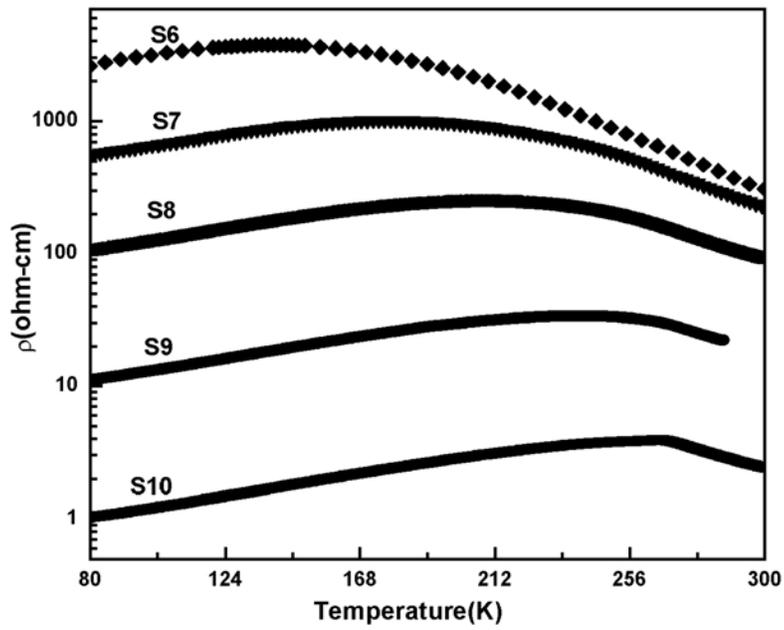

**Figure 5**

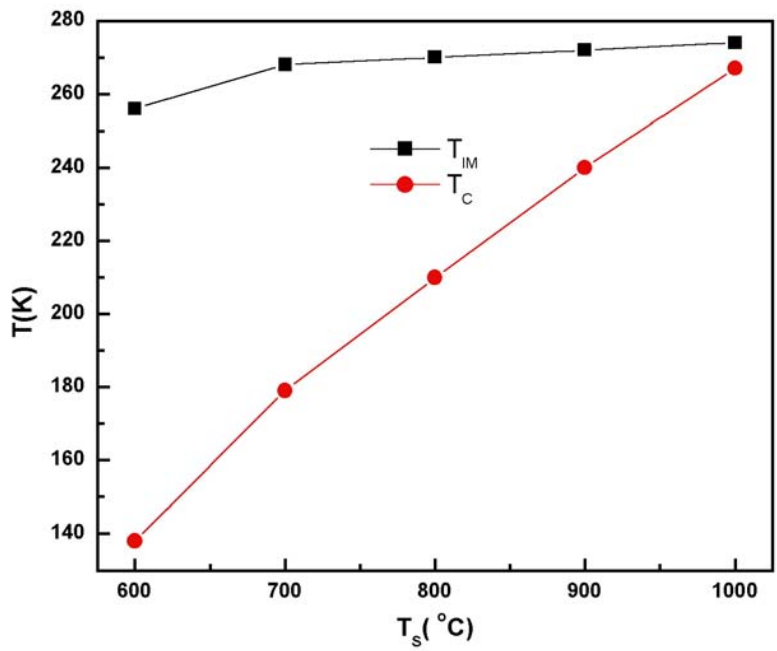

**Figure 6**

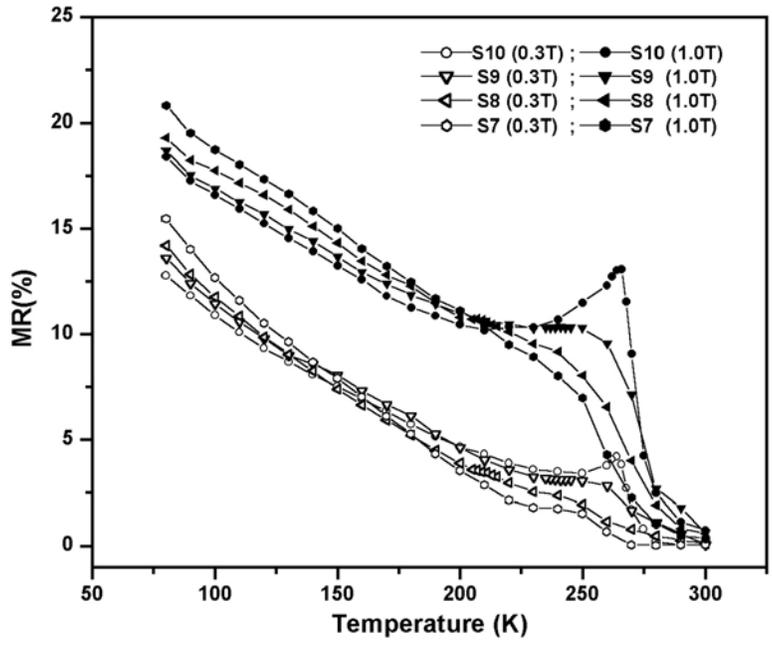

**Figure 7**

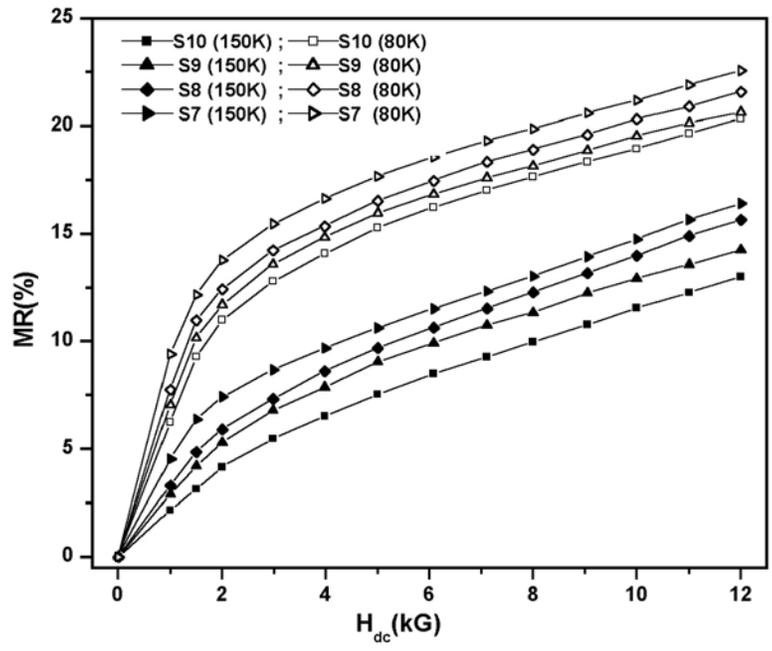

**Figure 8**

**Table-1**

| Sintering Temp.(°C) Ts | Lattice parameters | | | Cell volume(Å³) | Crystallite size(nm) XRD | Particle size(nm) SEM |
|---|---|---|---|---|---|---|
| | a(Å) | b(Å) | c(Å) | | | |
| 600 | 5.468 | 5.530 | 7.839 | 237.082 | ~29 | ~35 |
| 700 | 5.465 | 5.512 | 7.854 | 236.934 | ~32 | ~50 |
| 800 | 5.460 | 5.494 | 7.895 | 236.876 | ~35 | ~75 |
| 900 | 5.474 | 5.501 | 7.784 | 234.404 | ~40 | ~125 |
| 1000 | 5.473 | 5.494 | 7.762 | 233.420 | ~62 | ~200 |

**Table-2**

| Sintering Temp.(°C) Ts | $T_C$(K) | $T_{IM}$(K) | LFMR (%) at 1.0 T | |
|---|---|---|---|---|
| | | | 80K | 150K |
| 600 | 256 | 138 | -- | -- |
| 700 | 268 | 179 | 21.19 | 14.75 |
| 800 | 270 | 210 | 20.32 | 13.99 |
| 900 | 272 | 240 | 19.54 | 12.90 |
| 1000 | 274 | 267 | 18.94 | 11.54 |